\documentclass[prl,twocolumn,amssymb]{revtex4}
\usepackage{graphicx}% Include figure files
\usepackage{dcolumn}% Align table columns on decimal point
\usepackage{bm}% bold math
\usepackage{wasysym}
\usepackage{epstopdf}
\newcommand{\beq}{\begin{equation}}
\newcommand{\eeq}{\end{equation}}
\newcommand{\beqn}{\begin{eqnarray}}
\newcommand{\eeqn}{\end{eqnarray}}

\begin{document}
\title{Ising and Spin orders in Iron-based Superconductors}

\author{Cenke Xu}
\affiliation{Department of Physics, Harvard University, Cambridge
MA 02138}

\author{Markus M{\"u}ller}
\affiliation{Department of Physics, Harvard University, Cambridge
MA 02138}

\author{Subir Sachdev}
\affiliation{Department of Physics, Harvard University, Cambridge
MA 02138}

\date{\today}

\begin{abstract}

Motivated by recent neutron scattering experiments, we study the
ordering of spins in the iron-based superconductors
$\mathrm{La(O_{1-x}F_x)FeAs}$, assuming them in proximity to a
Mott insulator in the phase diagram. The ground state of the
parent system with $x = 0$ is a spin density wave with ordering
wave vector $\vec{Q} = (0 , \pi)$ or $(\pi, 0)$. Upon raising the
temperature, we find the system to restore SU(2) symmetry, while an
Ising symmetry remains broken, explaining the experimentally
observed lattice distortion to a monoclinic crystal structure.
Upon further temperature increase, the spins finally disorder at a
second transition. The phase transition driven by doping with
charge carriers similarly splits into an O(3) transition, and an
Ising transition with $z = 3$ at larger doping.

\end{abstract}
 \maketitle

\begin{figure}
\includegraphics[width=3.2in]{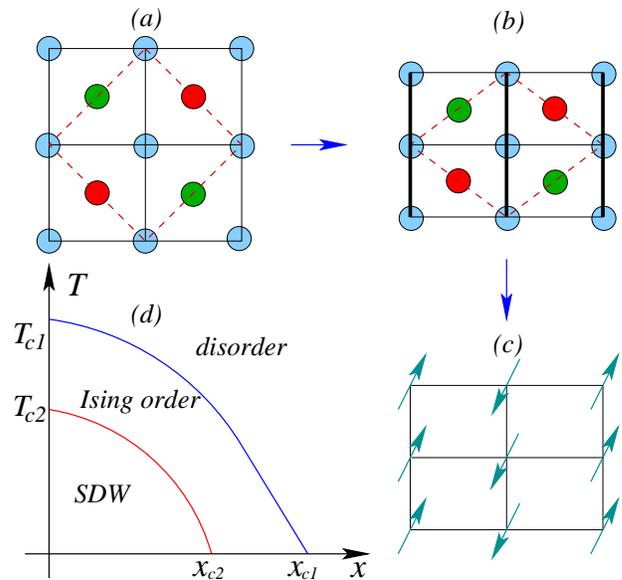}
\caption{$(a)$, the lattice structure at room temperature. The
gray circles are Fe ions, the green and red circles are As ions
above and below the Fe plane, respectively. The dashed square is
the two-Fe unit cell. $(b)$, the lattice structure between 150K
and 134K after the Ising order is developed: the thick lines
represent the bonds between antiparallel aligned spins, but no
uniform spin order is formed. The one-Fe unit cell is
orthorhombic, but the two-Fe unit cell has a 3d monoclinic
structure, as was seen in Ref. \cite{neutron2008}. $(c)$, the $(0,
\pi)$ spin order below 134K. $(d)$, the global phase diagram as a
function of temperature and doping $x$. The blue curve represents
the Ising transition, the red curve represents the O(3)
transition.} \label{phasedia}
\end{figure}

After more than two decades of prevailing in condensed matter
physics, copper-based high temperature superconductors have very
recently given in to their iron-based cousins
\cite{tc01,tc02,tc03,tc04,tc05,tc06,tc07}. The newly discovered
materials $\mathrm{M(O_{1-x}F_x)FeAs}$ with $\mathrm{M}$ being
rare earths such as $\mathrm{La},\mathrm{Sm}$, have similar
layered structure with stacked FeAs planes, sandwiched with La and
O. Transport measurements show that the ground state of the
undoped parent system is not an insulator, and LDA calculations
have identified both a small electron pocket and a hole pocket at
the Fermi level \cite{lda2008}. However, it has been argued that
the system is actually close to a Mott insulator, and a lot of
physics can be studied in a similar manner as the copper-based
high $T_c$ family, especially in the undoped system \cite{si2008}. In
our current work, we will %take the perspective of the Mott insulator,
%and
study the magnetism of these materials. Although the true unit
cell of the FeAs plane contains two Fe ions, because of the
staggered out-of-plane distribution of As ions (Fig.
\ref{phasedia}), we are only interested in the magnetic Fe
ions and so will use a unit cell with one Fe ion,
unless stated otherwise.

Recent neutron scattering experiments have shown that by lowering
temperature the undoped material first undergoes a structural
phase transition at 150K, with a distortion from tetragonal
structure to monoclinic structure, followed by a spin ordering
phase transition at 134K developing stripe order at $(\pi, 0)$
\cite{neutron2008}. The observed lattice distortion and spin
density wave (SDW) pattern are depicted in Fig.~\ref{phasedia}. In
the superconducting material with $x =0.08$, the SDW order is not
observed and, surprisingly, the distortion is absent as well,
which suggests that the lattice distortion is driven by the
development of the spin order. Interestingly, however, the SDW
order at low doping only appears at a lower temperature (134K)
than the lattice distortion (150K). We argue here that the SDW
ordering is preceded, both in temperature and doping, by the
breaking of an Ising symmetry in the effective spin model, and
that this is responsible for the observed lattice distortion.
%which however preserves spin
%rotation symmetry
%and is hard to detect directly above the SDW
%transition.

Ref.~\cite{si2008} argued that the undoped material is described
by either a $S = 1$ or $S = 2$ spin model with nearest and
next-nearest neighbor couplings $J_1,J_2$ that depend on the
competition between the onsite Hubbard interaction and the Hunds
rule: \beqn H = \sum_{<i,j>}J_1\vec{S}_i\cdot\vec{S}_j + \sum_{\ll
i,j \gg} J_2 \vec{S}_i\cdot \vec{S}_j. \label{j1j2} \eeqn Upon
doping, this $J_1$-$J_2$ model has $d_{x^2-y^2} + i d_{xy}$ and
then $d_{xy}$ superconductivity as $J_2$ increases into the regime
where the insulator has $(\pi, 0)$ SDW order \cite{sachdev02}.
%The LDA computation
%instead suggests the system has spin-1 or spin-3/2
%\cite{xiang2008}.
There is also a much weaker interlayer coupling $J_{\perp}$, which
is necessary to stabilize the spin order. It was suggested by
first principle calculations that both $J_1$ and $J_2$ are large
and antiferromagnetic \cite{yildirim2008}, and Ref.
\cite{xiang2008} showed that $J_2 \sim 2J_1$. It is well-known
that when $J_1 < 2J_2$, the classical ground state manifold of
model (\ref{j1j2}) is $S^2\otimes S^2$, because the two
sublattices of the square lattice will each form a N\'eel order
($\vec{n}_1$ and $\vec{n}_2$), and the ground state energy is
independent of the relative angle between these two N\'eel
vectors. However, quantum or thermal fluctuations lift the
degeneracy, leading to parallel or antiparallel alignment of the
two sublattice N\'eel vectors
\cite{henley1989,coleman1990,sachdev1991a}. If we define O(3)
vectors $\vec{\phi}_i$ as $\vec{n}_i$ with softened unit-length
constraint, the long wave-length field theory %for thermal
%fluctuations of the system is
reads: \beqn L &=& \sum_{a = 1}^2\sum_{\mu = x,y}
\partial_\mu\vec{\phi}_a\cdot \partial_\mu\vec{\phi}_a - r\vec{\phi}_a^2 +
u (\vec{\phi}_a^2)^2  + L^\prime, \cr\cr L^\prime &=& \gamma
\vec{\phi}_1
\partial_x\partial_y \cdot \vec{\phi}_2 - \alpha
(\vec{\phi}_1\cdot \vec{\phi}_2)^2.  \label{field} \eeqn In the
above equation we have absorbed the overall energy scale into
$\vec{\phi}_a$. The parameter $r$ is tuned by temperature, $\gamma
\sim J_1 / J_2$, $\alpha$ has contributions from both quantum and
thermal fluctuations: $\alpha \sim J_1^2/J_2^2\times (S \gamma_Q +
\gamma_T T/J_2)$, coefficients $\gamma_Q$ and $\gamma_T$ are given
in Ref. \cite{coleman1990}. $L'$ contains all sublattice couplings
preserving the square lattice symmetry. The latter rules out the
term $\vec{\phi}_1\cdot \vec{\phi}_2$, but allows for the coupling
$(\vec{\phi}_1\cdot\vec{\phi}_2)^2$.

The ground state manifold of the field theory (\ref{field}) is
$S^2 \otimes Z_2$, and the $Z_2$ order can be described by the
Hubbard-Stratonovich field $\Phi$, which couples to
$\vec{\phi}_1\cdot \vec{\phi}_2$: $ L^\prime = - \Phi
(\vec{\phi}_1 \cdot \vec{\phi}_2) + \Phi^2/(4\alpha)$. The ordered
state with $\Phi = 1$ ($\Phi = -1$) corresponds to the $(\pi , 0)$
($(0, \pi)$) SDW order. States with Ising $\Phi$ order, but only
short-range SDW order first appeared
\cite{sachdev1989,sachdev1991a} in the quantum theory of $H$ for
$S = 1, 2$. If the coupling $\alpha$ is relevant, an Ising
variable $\sigma$ can be introduced directly as $\vec{\phi}_2 =
\sigma\vec{\phi}_1$. Ref.~\cite{coleman1990} showed that thermal
fluctuations renormalize the anisotropy mixing $\gamma$ to zero at
long wavelength, so that at large scales the Lagrangian
(\ref{field}) can also be viewed as the low energy field theory of
the following Ising-O(3) model on the square lattice: $ H =
\sum_{<i,j>}J (1 + \sigma_i\sigma_j)\vec{n}_i \cdot \vec{n}_j$.
The O(3) vector $\vec{n}$ denotes either of $\vec{n}_1$ or
$\vec{n}_2$, and the coarse-grained mode of $\sigma$ is precisely
the Ising field $\Phi$ introduced before. The easy-plane version
of the Ising-O(3) model, dubbed the Ising-XY model, has been used
widely as an effective model for the fully frustrated XY model on
the square lattice and the triangular lattice
\cite{lee1984,lee1984a,kosterlitz1989,kosterlitz1991,kosterlitz1991a,kosterlitz1997,olsson1995}.

Note that the Ising order $\Phi\neq 0$ does not imply O(3) order;
however, because of the system is invariant under exchanging
$\vec{\phi}_1$ and $\vec{\phi}_2$, an O(3) order in
$\vec{\phi}_{1,2}$ implies Ising order. Therefore the transition
temperature of the Ising order is necessarily higher than that of
the O(3) symmetry breaking. If we consider a purely two
dimensional system, at finite temperature there is only a 2d Ising
transition separating an Ising ordered phase and a disordered
phase since a uniform O(3) order cannot exist at finite
temperature in dimensions smaller than 3. The transition
temperature can be estimated roughly as $T_{c1} / (J_2 \alpha )
\sim \xi^2/a^2 $, $\xi$ is the correlation length of the 2d O(3)
order at the transition, and $\xi^2/a^2$ is a factor gained from
integrating out the O(3) order parameters. A more precise estimate
of the Ising transition temperature for the $J_1-J_2$ model can be
found in Ref. \cite{coleman1990}, with $T_{c1}$ given by $ T =
0.13 \frac{J_1^2 S}{J_2} \times \frac{\xi(T)^2}{a^2}, \label{eq} $
in the large-$S$ limit. The Ising order breaks the $\pi/2$
rotation symmetry of the square lattice, indeed, an order
parameter $\Phi = 1$ implies that the spins tend to be aligned
parallel along $x$ but antiparallel along $y$. This Ising order
favors a lattice contraction in $y$ direction, {\em  i.e.\/},
towards the orthorhombic structure in Fig.~\ref{phasedia}. The
lattice distortion thus exists even in the absence of a uniform
O(3) order, but it necessarily requires the Ising order. A similar
mechanism was proposed for the lattice distortion in the cuprates
\cite{kivelson2006}.

%The physical idea is the same as given above.

The interlayer coupling $J_{\perp}$ will drive the 2d Ising
transition to a 3d Ising transition, but since it is much weaker
than the intralayer couplings, it will not move the transition
temperature significantly. However, the interlayer coupling
stabilizes an O(3) ordered phase at finite temperature, assuming
the interlayer coupling is small, the transition temperature can
be estimated as follows: The correlation length of the 2d O(3)
nonlinear sigma model scales as $\xi/a \sim \exp(2\pi/g)$, with $g
\sim \frac{T}{J_2m^2}$, $m$ is the magnetic moment of the SDW
order observed at zero temperature in units of the Bohr magneton,
which is empirically found to be only $m \sim 0.36$; the
interlayer coupling $J_{\perp}$ grows under renormalization, and
becomes unperturbative when $ J_{\perp}/(J_2m^2) \times \xi^2/a^2
\sim 1$, which will lead to the transition temperature $T_{c2}\sim
4\pi J_2 m^2/ \ln (m^2J_{2}/J_\perp)$ \cite{note}. $J_2$ is
evaluated to be $\sim$1000K in Ref.~\cite{xiang2008}, using the
transition temperature from Ref.~\cite{neutron2008}, the
interlayer coupling $J_\perp$ is estimated to be of order
$10^{-4}J_2m^2$, which can be neglected as compared with other
interactions, unless we are very close to a critical point. The
small value of the moment $m$ is probably due to quantum
fluctuations at zero temperature, since the system can be close to
quantum phase transitions. Close to, but above $T_c$, the
correlation length of the system scales like in the 3d O(3)
universality class, but once $\xi_z/a \sim
[(T-T_{c2})/J_\perp]^{-\nu}$ shrinks to 1, the system crosses over
to two dimensional critical behavior. The fact that the lattice
distortion observed in experiments \cite{neutron2008} occurs at a
temperature which is relatively small compared to the exchange
interaction $J_2$ \cite{xiang2008} is probably due to the
proximity to quantum phase transitions, which is consistent with
the small magnetic moment observed at low temperature
\cite{neutron2008}. The phase diagram is depicted in Fig.
\ref{phasedia}. Notice that the lattice is compressed along the
ferromagnetic order direction of the SDW, as observed
experimentally \cite{zhao2008}.

%Using the solution of the RG equation for the O(3) nonlinear sigma
%model in 2d, the correlation length at temperature $T$ reads $\xi
%/ a \sim \exp(2\pi/g)$. According to Ref. \cite{xiang2008}, if the
%system is $S = 1$, $J_1 =$ 660K, $J_2 = $1360K, using equation
%(\ref{eq}) the critical temperature for the Ising transition is
%evaluated to be $ T_{c1} \sim 200\mathrm{K} $ which qualitatively
%agrees with the critical temperature for the lattice distortion
%observed in experiments \cite{neutron2008}.

\begin{figure}
\includegraphics[width=3.0in]{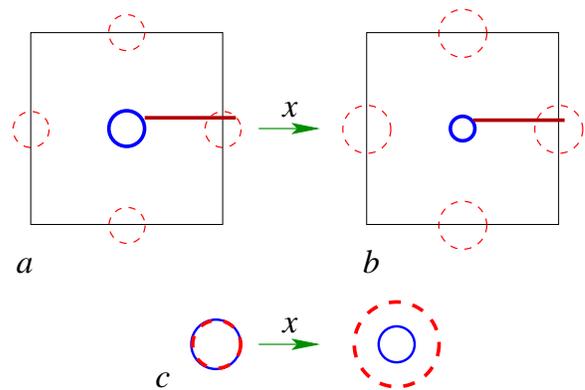}
\caption{$(a)$, the Brillouin zone for one-Fe unit cell at zero
doping. The thick blue circle denotes two almost overlaping hole
pockets \cite{mazin2008}, the red dashed circle represents the
electron pocket. The line which connects the electron and hole
pockets is the $(\pi, 0)$ wave vector. $(b)$, at finite doping,
the electron pocket expands and the hole pockets shrink, so that
the $(\pi, 0)$ vector can no longer connect points at the Fermi
level. $(c)$, translation of the electron pocket by vector $(\pi,
0)$: If at zero doping there is a perfect overlap of the pockets,
at infinitesimal doping there is no crossing between electron and
hole Fermi level at all, i.e. an order with $(\pi, 0)$ wave vector
cannot decay into particle-hole pair.} \label{nesting2}
\end{figure}

{\it Quantum phase transitions - } In $\mathrm{LaO_{1-x}F_xFeAs}$,
the SDW order vanishes as a small amount of extra carriers are
introduced by doping, meanwhile the superconductor state emerges,
implying the presence of one or more quantum phase transitions as
a function of doping. A tentative quantum critical point in these
systems has already been studied experimentally in a series of
samples $\mathrm{SmO_{1-x}F_xFeAs}$ \cite{qcp2008}. Since the
nature of the superconductor is not yet clear, however, in the
present work we focus on the quantum phase transitions of the spin
system discarding the presence of superconductivity. %If understood
%in terms of itinerant fermions, the SDW at $(\pi, 0)$ can be
%viewed as a result of the strong nesting between the electron and
%hole pockets (Fig. \ref{nesting}). If extra electrons are doped
%into this system, the nesting is lost very rapidly because of the
%unequal size of the electron and hole pockets. The nesting being
%perfect only at $x = 0$, the SDW order parameter cannot decay into
%a particle-hole pair excitations preserving both momentum and
%energy, because the SDW wave vector $(\pi, 0)$ does not connect
%two pairs of points on the Fermi surface for finite $x$ (Fig.
%\ref{nesting}).
In terms of the itinerant fermions, the SDW at $(\pi, 0)$ can be
understood from the large susceptibility arising from the location
of electron and hole pockets in the Brillouin zone: there are low
energy electron-hole pair excitations at the $(\pi, 0)$ wavevector
(Fig~\ref{nesting2}). As extra electrons are doped into this
system, these low energy excitations disappear rapidly because of
the unequal sizes of the electron and hole pockets. The SDW order
parameter cannot decay into a particle-hole pair excitations
preserving both momentum and energy, because the SDW wave vector
$(\pi, 0)$ does not connect two pairs of points on the Fermi
surface for finite $x$ (Fig. \ref{nesting2}). After integrating
out electrons we would obtain the following $z = 1$ Lagrangian:
%for
%$\vec{\phi}_1$ and $\vec{\phi}_2$
\beqn L &=& \sum_{i =
1}^2\sum_{\mu = \tau, x, y}
\partial_\mu\vec{\phi}_i\cdot \partial_\mu\vec{\phi}_i - r\vec{\phi}_i^2 +
u |\vec{\phi}_i|^4  + L^\prime, \cr\cr L^\prime &=&
\gamma\vec{\phi}_1
\partial_x\partial_y \cdot \vec{\phi}_2 + \gamma_1 |\vec{\phi}_1|^2|\vec{\phi}_2|^2- \alpha
(\vec{\phi}_1\cdot \vec{\phi}_2)^2 , \label{field2} \eeqn which
contains no dissipative term. The first three terms of the
Lagrangian describe the two copies of 3d O(3) systems on the two
sublattices. %The instantons of $\vec{\phi}_1$ and $\vec{\phi}_2$
%are confined by the $\alpha$ term, and the composite carries
%trivial Berry phase for spin-1 and spin-2 cases
%\cite{sachdev1991,sachdev1991a}.
The first term in $L^\prime$ mixes $\vec{\phi}_1$ and
$\vec{\phi}_2$, and its scaling dimension is \beqn \Delta[\gamma]
= D - (2 + D - 2 + \eta) = - \eta. \eeqn $\eta = 0.0375$
\cite{vicari2003} is the anomalous dimension of $\vec{\phi}$ at
the 3d O(3) universality class, therefore the $\gamma$ term is
irrelevant. The second term in $L^\prime$ is allowed by symmetry
and hence will be generated under renormalization. Its scaling
dimension can be evaluated as \beqn \Delta[\gamma_1] = D -
2\Delta[|\vec{\phi}|^2] =  D - 2(D - \frac{1}{\nu}) =
\frac{2}{\nu} - D, \eeqn with the correlation length exponent $\nu
= 0.71$ for the 3d O(3) transition. The $\gamma_1$ term is thus
also irrelevant. However, the $\alpha$ term is relevant at the 3d
O(3) transition, since it has positive scaling dimension
$\Delta[\alpha] = 0.581$ \cite{vicari2003}. We expect this term to
split the two coinciding O(3) transitions into two transitions, an
O(3) transition and an Ising transition, as was found in the
Schwinger boson theory \cite{sachdev1991a,sachdev1991}. Again,
because the O(3) order of $\vec{\phi}_{1,2}$ implies Ising order,
the latter should occur after the O(3) transition, i.e. at larger
$x$. The distance in doping between the two transitions can be
estimated by scaling, ignoring possible higher order singular
perturbations mediated by electrons: \beqn \Delta x \sim
\frac{\Delta r_c}{r_c} \sim \alpha^{\frac{1}{\nu\Delta[\alpha]}} =
\left(\frac{J_1}{J_2}\right)^{\frac{2}{\nu\Delta[\alpha]}}. \eeqn
Note that the monopoles of $\vec{\phi}_1$ and $\vec{\phi}_2$ are
confined by the $\alpha$ term. The Berry phase for monopoles of
spin-$S$ system on the square lattice is proportional to $i \pi
S$, the monopole-composite of $\phi_1$ and $\phi_2$ carries a
trivial Berry phase for $S = 1$ and $S = 2$ cases
\cite{sachdev1991,sachdev1991a}, and hence is ignored hereafter.

Note that while the O(3) SDW order parameters $\vec{\phi}_1$ and
$\vec{\phi}_2$ cannot decay into particle-hole excitations since
the wave vector $(\pi, 0)$ does not connect pairs of points at the
Fermi level, the same is not true for the Ising order parameter
$\Phi \sim \vec{\phi}_1\cdot\vec{\phi}_2$ which orders at $(0,0)$.
Also since $\Phi$ changes sign under a $\pi/2$ rotation and
reflection along the axis $x = y$, but does not break any other
symmetry, $\Phi$ couples to the two-body $d-$wave density $\Phi_q
\sim \sum_k \mathrm{Sign}[k_x^2 - k_y^2]c^\dagger_{k + q/2}
c_{k-q/2}$, and hence can decay into particle-hole excitations.
The decay rate can be calculated using Fermi's Golden rule: \beqn
\mathrm{Im}[\chi(\omega, q)] &\sim& \int \frac{d^2k}{(2\pi)^2}
[f(\epsilon_{k+q}) - f(\epsilon_{k})]\delta(\omega -
\epsilon_{k+q} + \epsilon_k)\cr\cr &\times& |\langle k| \Phi_q
|k+q \rangle|^2 \ \sim \ c_0 \frac{\omega}{q}. \label{damp} \eeqn
The standard Hertz-Millis \cite{hertz1976} formalism leads to a
$z=3$ theory with Lagrangian \beqn L =
\Phi_{-q}(\frac{|\omega|}{c_0q} + c_1 q^2 + r)\Phi_q + \cdots
\label{field3} \eeqn The ellipses stand for all the quartic and
higher order terms of $\Phi$ which are irrelevant at this Gaussian
fixed point described by (\ref{field3}). Quadratic terms with
singular factor $\omega^2/q^2$ or higher may occur in the
expansion, but since the theory has $z = 3$ these terms are
irrelevant. The $z = 3$ critical field theory was also obtained
for the electronic nematic phase with an order parameter similar
to $\Phi$ \cite{fradkin2001}. The critical exponents can be
extracted directly from the field theory (\ref{field3}). For
instance, in the quantum critical region, the specific heat and
the critical temperature of the finite temperature Ising
transition scale as \beqn C_v &\sim& T^{d/z} = T^{2/3}, \cr T_{c1}
&\sim& (x_{c1}-x)^{z /(d - 2 + z)} = x_{c1}-x,  \eeqn with $d =
2$. The weak interlayer coupling $w\Phi_{n}\Phi_{n+1}$ will
finally drive the scaling back to three dimensional behavior with
$w \sim J_{\perp}/J_2$, but its role is not considerable unless
the 2d correlation length is long enough, i.e., if we are close
enough to the quantum critical point. The spatial scaling
dimension of $w$ is $\Delta[w] = 2$ at the 2d critical point
described by Eq. \ref{field3}, therefore $w$ becomes
nonperturbative when $(\xi / a)^{\Delta[w]} \sim 1/w$, i.e. \beqn
x_{c1}-x  \sim w^{1/(\nu \Delta[w])} = w. \eeqn Within this small
window, the critical scaling becomes \beqn C_v &\sim& T^{d/z} = T,
\cr T_c &\sim& (x_{c1}-x)^{z /(d - 2 + z)} = (x_{c1}-x)^{3/4}.
\eeqn

The O(3) order parameter $\vec{\phi}$, which can be taken as
$\vec{\phi}_1$, cannot decay into particle-hole pairs, assuming
the $(\pi, 0)$ wave vector does not connect two points at the
Fermi level. The Gaussian part of the Lagrangian describing
the O(3) transition at $x_{c2}$ has dynamical exponent $z = 1$: \beqn L =
\vec{\phi}_{-q}(\omega^2 + q^2) \vec{\phi}_q + L^\prime.
\label{field4}\eeqn $L^\prime$ consists of quartic and higher
order terms. If the quartic terms have no singularity in momentum
and frequency space, the Lagrangian (\ref{field4}) describes a 3d
O(3) transition. Berry phases of monopoles in this case are
trivial for spin-1,2 \cite{sachdev1991a,sachdev1991} and so are
not noted. However, the quartic terms of the effective action may
include singular terms like \beqn L^\prime = \gamma_2
|\vec{\phi}|^2_{-q}\frac{|\omega|}{q}|\vec{\phi}|^2_q. \eeqn This
term can be viewed as describing the decay of $|\vec{\phi}|^2$,
which couples to the zero momentum charge density. From naive
power-counting, $\gamma_2$ has the same scaling dimension as all
the other quartic terms without singularities. However, since it
mixes the $|\vec{\phi}|^2$ field at distinct spatial points, the
anomalous dimensions will be contributed by the two different
points separately. Therefore its scaling dimension can be
evaluated as $ \Delta[\gamma_2] = D - 2\left(D -
\frac{1}{\nu}\right) = \frac{2}{\nu} - D$,  which is again
irrelevant at the 3d O(3) transition. If no other more relevant
quartic terms are present, the quantum phase transition of
$\vec{\phi}$ at $x_{c2}$ belongs to the 3d O(3) universality
class, cf.~Fig.~\ref{phasedia}. But a thorough analysis of the
quartic terms is required to draw a firm conclusion.

Quantum critical points play an important role in transport because
the electrons can scatter off the critical modes. We expect the
Ising critical modes to contribute the dominant part to the low
temperature resistivity, because of its $z = 3$ soft modes and the
ensuing larger density of states at low energy. At low temperature
where the scattering is dominated by small angle forward scattering, the
resistivity is expected to scale as $\rho \sim T^{4/3}$. The more
general formula for the resistivity for a $z = 3$ theory with Lagrangian
(\ref{field3}) reads $\rho \sim T^{(d+2)/z}$, which is consistent
with the well-known $T^{5/3}$ law of the resistivity at the
quantum critical point of three dimensional itinerant
ferromagnetic order \cite{ueda1975}.

%; at intermediate temperature (but still in quantum critical
%region) the scattering rate is isotropic, and the resistivity
%would scale as $\rho \sim T^{2/3}$. In Ref. \cite{qcp2008}, above
%the superconducting transition temperature, at the vicinity of the
%quantum critical point the resistivity seems to scale with the
%temperature with a power smaller than 1, which is consistent with
%our calculation.

In summary, we have studied the SDW at $(\pi, 0)$ observed
experimentally in $\mathrm{LaO_{1-x}F_xFeAs}$, and its phase
transitions. While raising the thermal and quantum fluctuations,
the SDW is predicted to cede to a state with restored SU(2)
invariance, but retaining a broken Ising symmetry which drives a lattice
distortion. This is followed by an Ising transition at higher
temperature or larger doping. The nature and universality classes
of these transitions, and various critical exponents are discussed.
\\
{\em Note added:} Fang {\em et al.} \cite{kivelson2008} have also
applied thermal fluctuations of the $J_1$-$J_2$ model to the
iron-based superconductors.

\bibliography{feas}

\end{document}